\documentclass[5p]{elsarticle}

\usepackage[T1]{fontenc}
\usepackage[english]{babel}
\usepackage{natbib}
\usepackage{graphicx,subfigure}
\usepackage{amsfonts,amssymb, dsfont}
\usepackage[cmex10]{amsmath}
\usepackage{algpseudocode,algorithmicx,algorithm}

\usepackage{fixltx2e}
\usepackage{color}
\usepackage{verbatim}
\usepackage{subfloat}
\usepackage{enumitem}
\usepackage{tikz}
\usetikzlibrary{positioning, decorations.text}

\def\@{\hskip.8pt}
\def\?{\hskip.3pt}
\def\plus#1#2{\vrule height#1pt width0pt depth#2pt}

\def\mbf#1{\mathbf{#1}}
\def\narc#1{\text{\@ \scriptsize $(#1)$}}
\def\nard#1{\text{\@ \tiny $(#1)$}}

\def \red#1 {\textcolor{red}{#1}}
\def \blue#1 {\textcolor{blue}{#1}}

\begin{document}

\begin{frontmatter}

\title{Bayesian Multi--Dipole Modeling in the Frequency Domain}
\author[dima]{Gianvittorio Luria}
\author[besta]{Dunja Duran}
\author[besta]{Elisa Visani}
\author[aalto]{Sara Sommariva}
\author[besta]{Fabio Rotondi}
\author[besta]{Davide Rossi Sebastiano}
\author[besta]{Ferruccio Panzica}
\author[dima,cnr]{Michele Piana} 
\author[dima,cnr]{Alberto Sorrentino\corref{cor1}}
\ead{sorrentino@dima.unige.it}
\cortext[cor1]{Corresponding author. Address: Department of Mathematics, University of Genoa, 
               Via Dodecaneso $35$,  $16146$ Genova, Italy.\\ Tel.:\,\,$+39\ 0103536644$.}
\address[dima]{Department of Mathematics, University of Genoa, Genoa, Italy}
\address[cnr]{CNR - SPIN, Genoa, Italy}%
\address[besta]{Department of Neurophysiology and Diagnostic Epileptology, IRCCS Foundation Carlo Besta Neurological Institute, Milan, Italy}
\address[aalto]{Department of Neuroscience and Biomedical Engineering, Aalto University School of Science, Espoo, Finland}

\begin{abstract}
\emph{Background\@:} Magneto- and Electro--encephalography record the electromagnetic field generated by neural currents with high temporal frequency and good spatial resolution, and are 
therefore well suited for source localization in the time and in the frequency domain. In particular, localization of the generators of neural oscillations 
is very important in the study of cognitive processes in the healthy and in the pathological brain.

\noindent \emph{New method\@:} We introduce the use of a Bayesian multi--dipole localization method in the frequency domain. Given the Fourier Transform of the data at one or multiple frequencies and/or trials, the algorithm approximates numerically the posterior distribution with Monte Carlo techniques.

\noindent \emph{Results\@:} We use synthetic data to show that the proposed method behaves well under a wide range of experimental conditions, including low signal--to--noise ratios and correlated sources. We use dipole clusters to mimic the effect of extended sources. In addition, we test the algorithm on real MEG data to confirm its feasibility.

\noindent \emph{Comparison with existing method(s)\@:} Throughout the whole study, DICS (Dynamic Imaging of Coherent Sources) is used systematically as a benchmark. The two methods provide similar general pictures; the posterior distributions of the Bayesian approach contain much richer information at the price of a higher computational cost.

\noindent \emph{Conclusions\@:} The Bayesian method described in this paper represents a reliable approach for localization of multiple dipoles in the frequency domain. 
\end{abstract}


\end{frontmatter}

\bibliographystyle{elsarticle-num} 

\section{Introduction}\label{S0} 

Magneto-- and Electro--Encephalography (M/EEG) 
\linebreak
stand out among the functional neuroimaging techniques for the temporal resolution of their recordings, of the order 
of the millisecond \cite{hamalainen1993magnetoencephalography,del2001magnetoencephalography,hansen2010meg}. 
In principle, this feature makes M/EEG perfect tools in order to investigate the precise timing of brain responses 
to external stimuli \cite{pang2011localization, huang2010somatosensory,sorrentino2006modulation}, to disentangle the complex 
interactions of neural populations organized in connectivity networks \cite{de2010temporal, schoffelen2009source}, 
or to locate the onset of epileptogenic activity \cite{wilenius2013interictal, von2016detection,heers2014frequency}. 
In practice, however, one must cope with the fact that spatial mapping of brain activity from M/EEG data requires to solve an ill--posed inverse problem \cite{dassios2013definite}.

From a methodological perspective, the inverse M/EEG  problem is typically solved in the time domain. In the last fifteen years, most efforts have been devoted to improving the reliability of the source reconstructions by exploiting smoothness in the temporal 
domain \cite{li2011spatio,gramfort2012mixed,sommariva2014sequential,vivaldi2016bayesian}.
However, in all circumstances in which neurophysiological phenomena exhibit a repetitive/oscillatory nature, performing source modeling in the frequency domain might help in solving the inverse problem, by providing automatic noise filtering and 
a natural unmixing of the contributions of sources at different frequencies.
Noteworthy examples of oscillatory neurophysiological phenomena are brain rhythms \cite{da1977cortical,pfurtscheller1997existence,pineda2005functional} and  
resting state networks \cite{de2010temporal,bretal11}; in addition, recent experimental evidence suggests that high frequency oscillations might be considered to be biomarkers of epilepsy \cite{roehri2016time,cimbalnik2018cs}; finally, frequency representation of the signal is sometimes used in the development of brain--computer interfaces \cite{lotte2018review}.

To the best of our knowledge, only relatively few options are available for source localization in the frequency domain.
These include classical regularization with \linebreak
minimum--norm estimates (\/MNE\/) and minimum--current estimates (\/MCE\/) 
\cite{jensen2002new, liljestrom2005neuromagnetic}, wMEM \cite{lina2014wavelet}, a technique based on the maximum entropy on the mean
in the wavelet domain,  and Dynamic Imaging of Coherent Sources (\/DICS\/) \cite{gr01}, a frequency domain spatial filter 
employing a minimum variance adaptive beamformer approach. Single dipole fitting of complex Fourier transformed data was done in 
\cite{tesche1993comparison}; we are not aware of any attempt at using multi--dipole estimation methods in the frequency domain.

In this work, we extend the Semi--Analytic Sequential Monte Carlo sampler (\/SASMC\/) described in \cite{sommariva2014sequential} and show that it can be effectively used 
for estimating multiple dipoles from Fourier transformed data.

The SASMC method is a numerical technique implementing a Bayesian semi--analytic approach to conditionally linear inverse problems, 
namely to inverse problems in which the forward model establishes a linear dependence of the data on a subset of the unknowns.
The algorithm presented here takes in input one or more spatial distributions of Fourier transformed M/EEG data, each picked at a single frequency, 
and returns the approximation of the joint posterior probability distribution of the number of sources and of their parameters. 
Importantly, the number of sources needs not be set in advance, as it is automatically estimated from the data, and 
the Monte Carlo procedure is only applied to the nonlinear variables, while the marginal posterior distribution of the linear ones is 
computed analytically.
Moreover, thanks to the use of the Fourier transformed data, rather than of power spectrum, the algorithm exploits the information on the phase of the signal, which should theoretically improve localization. 
Lastly, unlike beamformers, the method is not based on data covariance matrix (\/or on cross spectral density\/) and is therefore less affected by intersource correlation \cite{belardinelli2012source}.

\smallskip
The plan of the paper is as follows. 
Section \ref{S2} first briefly outlines the multi--dipolar model framework for the neuromagnetic inverse problem, in which primary currents are modelled as the superposition of point--like dipoles applied in a discretized spatial grid. Then the key points of SASMC are summarized, 
and a schematic description of the algorithm steps is given.
In Section \ref{S3} source modeling in the frequency domain is performed by means of SASMC. In the first instance several syntethic datasets with different signal--to--noise ratios, intersource correlations and source extent are analyzed and subsequently the method is tested against experimental data. In all cases, results provided by DICS are also given as a touchstone.
Eventually, in Section \ref{S4}, a discussion of the result is presented and our conclusions are offered.

\section{Methods}\label{S2}
The next subsections provide an overview of multiple dipole modeling with SASMC.
Such method adopts a Bayesian perspective on the problem of estimating the 
parameters of an unknown number of current dipoles 
from a set of spatial distributions of complex electromagnetic field. 
For a detailed description, the reader is referred to \cite{sommariva2014sequential, soluar14}.

\subsection{Multi--dipole state--space.}\label{SS2.1} 
The neuronal activity producing the electromagnetic field measured by M/EEG is modelled using a primary current distribution which, in this work, is approximated by the superposition of an unknown number of current dipoles \cite{hamalainen1993magnetoencephalography,baillet01}.

In mathematical terms, a single dipole is represented by a pair $(r, \mbf{q})$, where $r$ is an integer variable representing the dipole location in a given discretized source space, and $\mbf{q}$ is a three--dimensional vector representing the dipole moment. The pair $(r, \mbf{q})$ can also be seen as a point in a corresponding single--dipole space $\mathcal{D}$. A couple of dipoles can therefore be seen as a point in the corresponding double--dipole space $\mathcal{D}^{\/\/2} = \mathcal{D}\times \mathcal{D}$, $\times$ denoting the Cartesian product; more generally, a $n_D$--tuple of dipoles is a point in $\mathcal{D}^{\/\/n_D}$. Since in our approach the number of dipoles is among the unknowns, the state--space of the unknown primary current $\mbf{j}\@$ is eventually defined as the disjoint union of 
spaces  \cite{soluar14}, \cite[p.~488]{camory05}:
\begin{equation}\label{eq:state_space}
 \mathcal{J}\ :=\ \bigcup_{n_D}\ \{n_D\} \times \mathcal{D}^{\@n_D}\@,
\end{equation}
with $\mathcal{D}^{\@0} := \{\emptyset\}$.
Any current distribution $\mbf{j}$ is therefore represented as
\begin{equation}\label{eq:j}
 \mbf{j} = \left(\/n_D,\@ r_1,\@ \mbf{q}_{\/1},\@ \ldots,\@ r_{n_D},\@ \mbf{q}_{\/n_D}\/\right)\ \in \ \mathcal{J}\ ,
\end{equation}
or also equivalently as
\begin{equation}\label{eq:j_2}
  \mbf{j} = \left(\/n_D,\@ \mbf{r}_{1:n_D},\@ \mbf{q}_{1:n_D}\/\right)\ ,
\end{equation}
which directly follows from \eqref{eq:j} by reordering the axes and by introducing the shorthand notations
\begin{align*}
 \mbf{r}_{1:n_D}\, &:=\, \left(\@r_1, \ldots, r_{n_D}\@\right)\@ ,\\
 \mbf{q}_{1:n_D}\,  &:=\, \left(\@\mbf{q}_{\/1}, \ldots, \mbf{q}_{\/{n_D}}\@\right)\@. 
\end{align*}

We shall henceforth stick to the representation \eqref{eq:j_2}.
%

\subsection{Statistical model.}\label{SS2.2} 
Let $\{\mbf{y}_t\}_{\/t=1, \ldots, T}$\vspace{1pt} be the time series of M/EEG recordings; each \vspace{1pt}$\mbf{y}_t = (\/{y_t}^{\!1}, \ldots, {y_t}^{\!n_s}\/)$ is an array, whose
$i$--th element is the measurement made at time $\/t\@$ by the $i$--th sensor.
We assume data to be affected by zero--mean Gaussian additive noise, so that at each sampled time $t$
\begin{equation}\label{eq:obs_time}
\mbf{y}_t = \mbf{e}_t + \mbf{n}_t\@,
\end{equation}
being $\mbf{e}_t$ the exact field induced by the primary current distribution $\mbf{j}_t$ and $\mbf{n}_t$ the noise term. 
The explicit model for $\mbf{e}_t$ is given by
\begin{equation}\label{eq:exact_model}
\mbf{e}_t = \sum_{k=1}^{n_D} G(r_k) \cdot \mbf{q}_{\@k,t} =:\@  G\!\left(\@\mbf{r}_{1:n_D}\/\right) \cdot \mbf{q}_{\@1:n_D,t}\@,
\end{equation}
where, at time $t\/$,  $G(r_k)$ is the lead field matrix computed at the location $r_k$ of the $k$--th dipole on the discretized cortex,  
$\mbf{q}_{\/k,t}$ is the corresponding dipole moment, and
\begin{align*}
 \mbf{q}_{\@1:n_D,t}\,  &:=\, \left(\@\mbf{q}_{\/1,t}, \ldots, \mbf{q}_{\/{n_D}, t}\@\right)\@ ,\\[1pt]
 G\!\left(\/\mbf{r}_{1:n_D}\/\right)\, &:=\, \bigg[\@G(r_1)\@\bigg\lvert\,\cdots\@\bigg\lvert\@G(r_{n_D})\@\bigg]\@ .
\end{align*}
In \eqref{eq:exact_model} it is assumed that the number of sources $n_D$ as well as their locations $r_1, \ldots, r_{n_D}$ do not
change with time. 

\smallskip
Denoting by $\{\hat{\mbf{y}}_f\}_{f=1,...,F}\@$  the  Discrete Fourier Transform (\/DFT\@) of  \vspace{1pt}$\{\mbf{y}_t\}$, the 
linearity of the Fourier operator entails that
a formula analogous to \eqref{eq:obs_time} holds in the frequency domain:
\begin{equation}\label{eq:obs_freq}
\hat{\mbf{y}}_f = \hat{\mbf{e}}_f + \hat{\mbf{n}}_f\, =:\, G\!\left(\@\mbf{r}_{1:n_D}\@\right) \cdot \hat{\mbf{q}}_{\@1:n_D,f} + \hat{\mbf{n}}_f\@,
\end{equation}
where the distribution of $\hat{\mbf{n}}_f$ is  still Gaussian and zero--mean.

\medskip 
From a mathematical point of view, the only difference between \eqref{eq:obs_time} and \eqref{eq:obs_freq} is that the former involves real quantities, 
while the latter involves complex ones.
Therefore, if \eqref{eq:obs_time} is seen as a system of equations in $\mathds{C}$, the analogy is complete. 
This means that the inverse problem of making inference on $\mbf{j}$ given a single topography $\mbf{y}$ can be considered in abstract terms, 
regardless of the fact that the latter represents the spatial distribution of the electromagnetic field at a single time point or the Fourier transform 
of the data picked at a single frequency. 
As a consequence the same machinery described in \cite{sommariva2014sequential} for Bayesian inference of 
multiple dipoles in the time domain can be used in the frequency domain. 
In particular, as shown in \cite{sommariva2014sequential}, this approach easily generalizes to include multiple topographies.
As we will show with both simulated and experimental data, this implies that we can use the method to estimate dipoles from data taken both at different 
frequencies and/or from different trials.

We now provide a brief overview of the methodology. In order to avoid a too complicated notation, the description below deals with inference from a single topography.

\smallskip
In the Bayesian approach to the problem \cite{evst02,soka04}, the data $\mbf{y}$, the unknown $\mbf{j}$ (as defined in equation \eqref{eq:j_2}) and the noise $\mbf{n}$ are considered as the realizations of corresponding random variables $\mbf{Y}$,
$\mbf{J} = \left(\@N_D,\, \mbf{R}_{1:n_D},\, \mbf{Q}_{1:n_D}  \,\right)$ and $\mbf{N}$, whose functional relation is given by
\begin{equation}\label{eq: obs_rv}
\mbf{Y}\, =\, F\/(\@\mbf{J}\@) + \mbf{N}\, =\, G\!\left(\@\mbf{R}_{1:n_D}\right)\@\cdot\@ \mbf{Q}_{1:n_D}\@ +\@ \mbf{N}\, .
\end{equation}

In this framework the solution is the posterior probability distribution of $\mbf{J}$ conditioned on the data, which, 
in the light of Bayes' theorem, can be written as
\begin{equation}\label{eq:posterior}
p(\@\mbf{j}\@|\@\mbf{y}\@) = \frac{p\/(\mbf{y}\@|\@\mbf{j}\@)\,p(\@\mbf{j}\@)}{p(\mbf{y})}\@.
\end{equation}
In \eqref{eq:posterior}, $p(\@\mbf{j}\@)$ is the prior probability distribution of $\@\mbf{J}\@$, encoding all the information on the unknown which is available before the measurement is made;
$p\/(\mbf{y}\@|\@\mbf{j}\@)$ is the likelihood function, containing information regarding the forward model and the statistical properties of the noise; and $p(\mbf{y})$ is a 
normalizing constant whose knowledge is not necessary for the analysis described below.
From $p(\@\mbf{j}\@|\@\mbf{y}\@)$ sensible estimates of $\@\mbf{j}$ can then be computed.

\smallskip
Given the definition \eqref{eq:state_space} of the state space $\mathcal{J}$, the prior distribution 
for the unknown set of dipoles  is built as
the product of a prior distribution for the number of dipoles $n_D$ and a prior distribution on the corresponding space $\mathcal{D}^{\@n_D}$\@:
\begin{equation}
p(\@\mbf{j}\@) =  p(\/n_D\/) \prod_{k=1}^{n_D}\, p(\/r_k\@|\@n_D, r_1, \ldots, r_{k-1}\/)\, p(\@\mbf{q}_{\@k}\/)\@,
\label{eq:prior}
\end{equation}
where 
$\mbf{q}_{\@k}$ is the $k$--th dipole's moment, and  $r_0 = \emptyset\@$.

We assume $p(n_D)$ in \eqref{eq:prior} to be a Poisson distribution with mean $\lambda$; in general, when working with multi--dipole models, one aims at explaining the measured data with a small number of sources which implies choosing a small value, such as $\lambda = 0.25$, to discourage larger models; however, as we will see in the simulations below, different choices are possible.
Given the number of sources, dipole moments are assumed to be independent from dipole locations. 
The prior distribution for the source locations is uniform in the brain, under the constraint that the $n_D$ dipoles be located at different points; this causes the prior distribution of $r_k$ to be conditioned on the locations of the previous dipoles in \eqref{eq:prior}.
The prior distribution $p(\@\mbf{q}_{\@k}\/)$ for each dipole moment is a trivariate normal distribution, with zero mean and diagonal variance matrix equal to
$\sigma_q^2\,\mbf{I}_{\plus70 3}\@$. The parameter $\sigma_q^2$ reflects the information on the dipole strength and it can be roughly estimated from the data and from the forward model. In the simulations below we will show how changes in $\sigma_q^2$ and $\lambda$ interact to produce different results.

As far as the likelihood function $p\/(\mbf{y}\@|\@\mbf{j}\@)$ is concerned, noise is assumed to be Gaussian with zero mean and diagonal covariance 
matrix $\Gamma_n = \sigma_n^2\@\mbf{I}_{\plus60 n_s}$.

\medskip
Equation \eqref{eq: obs_rv} shows that, for each realization $(\@n_D, \mbf{r}_{1:n_D}\@)$ of $(\@N_D, \mbf{R}_{1:n_D}\@)$,
the random vector $\mbf{Y}$ depends linearly on $\mbf{Q}_{1:n_D}$.
Therefore \cite{sommariva2014sequential, soka04}, assuming the mutual independence of $\mbf{J}$ and $\mbf{N}$ and
under the Gaussian assumptions made above about the prior density for the dipole moment and the noise model,
the marginal likelihood $p(\mbf{y}\,|\, n_D, \mbf{r}_{1:n_D})$ is a Gaussian density with zero mean and covariance 
\begin{equation}
 \Gamma_l = \sigma_q^2\,  G\/(\mbf{r}_{1:n_D})\, G\/\left(\mbf{r}_{1:n_D}\right)^T +\, \sigma_n^2\,\mbf{I}_{\plus60 n_s}\, .
\end{equation}
Under the previous assumptions, the conditional posterior $p(\mbf{q}_{1:n_D}\,|\,\mbf{y}, n_D, \mbf{r}_{1:n_D})$ is also normally distributed \cite[Theorem~3.7]{soka04} with mean
\begin{subequations}\label{eq:post_q}
 \begin{equation}
  \sigma^2_q\@ \, G\/(\mbf{r}_{1:n_D})^{T}\, {\Gamma_l}^{-1}\@ \mbf{y}
 \end{equation}
 and variance
 \begin{equation}
  \sigma_q^2\,\mathbf{I}_{\plus60 3\@n_D} -\, \sigma^4_q\, G\/(\mbf{r}_{1:n_D})^{T}\, {\Gamma_l}^{-1}\@G\/(\mbf{r}_{1:n_D})\, .
 \end{equation}
\end{subequations}


\subsection{Approximation of the posterior distribution.}\label{SS2.3}

In order to compute estimates of the primary currents from the posterior distribution, a numerical approximation of the latter is needed. 
Since the posterior $p(\@\mbf{j}\@|\@\mbf{y}\@)$ is potentially a highly complex function on a high--dimensional space,
we resort to Sequential Monte Carlo samplers \cite{dedoja06} that behave very efficiently in such cases.

In particular, the SASMC sampler described in \cite{sommariva2014sequential} and used in this study exploits the semi--linear structure of the model \eqref{eq: obs_rv} by approximating the posterior 
\begin{equation}
 p(\@\mbf{j}\@|\@\mbf{y})\, =\,  p(\mbf{q}_{1:n_D}\,|\,\mbf{y}, n_D, \mbf{r}_{1:n_D}\@)\ p(n_D, \mbf{r}_{1:n_D}\,|\,\mbf{y}\@)
\end{equation}
through a two--step algorithm\@: first, an Adaptive Sequential Monte Carlo sampler (\/ASMC\/), described in \cite{soluar14} 
and summarized below, is used to approximate the marginal posterior $p(n_D, \mbf{r}_{1:n_D}\,|\,\mbf{y}\@)$ of the number of dipoles and of their location; then, the mean and covariance matrix of the conditional posterior $p(\mbf{q}_{1:n_D}\,|\,\mbf{y}, n_D, \mbf{r}_{1:n_D}\@)$ of the dipole moments are analytically computed through formulas \!\! (\ref{eq:post_q}\@a,\@b).

\medskip
\subsubsection{Adaptive Sequential Monte Carlo samplers}

The general idea underlying Monte Carlo methods is to approximate a target probability distribution using a large set of samples, also called \textit{particles}; in our context, the target probability distribution is the posterior $p(n_D, \mbf{r}_{1:n_D}\,|\,\mbf{y}\@)$ and each sample is a candidate solution, i.e.\! the number of dipoles and the dipole locations.

One easy way to produce such set of samples is to draw them independently from a simple distribution, and possibly weight them to correctly approximate the target distribution (\@Importance Sampling, IS \cite{roca04}\@). An alternative approach is to start from a random candidate, perturb it randomly many times, and then approximate the target distribution with the collection of samples along the iterations (\@Markov Chain Monte Carlo, MCMC \cite{roca04}\@).

The main drawback of IS is that hitting at random a good solution is extremely unlikely; the main drawback of MCMC is that it is difficult to jump out of a local maximum. In the class of methods known as Sequential Monte Carlo samplers \cite{dedoja06}, these two techniques are combined\/:  multiple samples are independently drawn from a simple distribution, evolve following an MCMC scheme, and their weights are updated after every MCMC step; at times, samples having negligible weights are replaced by samples in the higher--probability region, so as to explore better these areas. Eventually, the target distribution is approximated by the weighted sample set obtained at the last iteration.\\

More formally, three main ideas underlie the Adaptive SMC sampler used in this work.

First, instead of trying to directly sample the posterior distribution, the latter is smoothly reached through a sequence of auxiliary distributions 
\begin{subequations}
\begin{equation}\label{eq:sequence}
\big\{\@p_n(n_D, \mbf{r}_{1:n_D}\@|\@\mbf{y})\@\big\}_{n = 1, \ldots, N}\,,
\end{equation}
being $N$ the number of iterations. For each $n$, the corresponding $p_n$ is defined as
\begin{equation}
p_n(n_D, \mbf{r}_{1:n_D}\@|\@\mbf{y}) \propto p(\mbf{y}\@|\@n_D, \mbf{r}_{1:n_D})^{\alpha(n)} \/p(n_D, \mbf{r}_{1:n_D})
\end{equation}
\end{subequations}
%
with $\alpha(1)=0$, $\alpha(N)=1$ and $\alpha(1) < \alpha(2) < \ldots < \alpha(N)$.
In this way, the first distribution $p_1$ is the prior distribution; for $n >1$,  $p_n$ is obtained as a combination of the prior and the likelihood distributions, the latter being weighted more with the iterations, so that the information content of the data is embodied gradually into the sequence of distributions; the last distribution of the sequence is the target posterior distribution.

Second, IS and MCMC techniques are combined to approximate sequentially each distribution of the sequence \eqref{eq:sequence} as the weighted 
particle set 
\[\big\{X_n^\narc{i\/} := \left(n_D, \mbf{r}_{1:n_D}\right)_n^{\narc{i}} , w_n^\narc{i\/}\big\}_{i=1,...,I}\@\]
where each particle contains all the parameters that are estimated through the proposed Monte Carlo procedure, namely the number of active sources and their location. 
The number of particles $I$ represents, roughly speaking, the number of candidate solutions that are tested in the Monte Carlo procedure; therefore the higher the number, the better the approximation, but at the price of a higher computational cost. 

Finally, the sequence of exponents $\{\alpha(n)\}_{\/n = 1, \ldots,  N}$ is not established a priori, but adaptively determined at run--time.
This means that the actual number of iterations is also determined online, even if it is always kept within given lower and upper bounds.\\

The algorithm works as follows.
At $n = 1$, the exponent $\alpha\/(1)$ is set to $0\@$; the initial sample set $\big\{X_1^\narc{i\/}\big\}_{i=1,...,I}$ is drawn from the prior distribution 
and assigned uniform weights $ w_1^\narc{i\/} = \frac{1}{I}\@\vspace{1pt}$. Subsequently, the following steps are iterated until $\alpha\/(n)$ reaches $1$:

\begin{itemize}
 \item 
  the sample set $\big\{X_{n}^\narc{i\/}\big\}_{i=1,...,I}$ is obtained from the previous one $\big\{X_{n-1}^\narc{i\/}\big\}_{i=1,...,I}\vspace{1.5pt}$ 
  by drawing each particle from  a $p_{n}$--invariant kernel, which is the product of a Reversible Jump Metropolis--Hastings kernel \cite{gr95}, accounting for a possible change 
  in the number of dipoles in the particle, and $n^\narc{i\/}_D$ Metropolis--Hastings kernels \cite{hastings1970monte}, for dipole locations evolution. This way, each particle explores 
  the state space by allowing both the number of dipoles as well as their locations to change.

The increment or decrement by one of $n^\narc{i\/}_D$ is attempted with probability of 
$\frac{1}{3}$ and of $\frac{1}{20}$, respectively. If a birth move is accepted, the location of the newborn dipole is uniformly distributed. 
If a death move is accepted, the excluded dipole is uniformly chosen among the existing ones.

As far as source locations are concerned, each dipole is let move only to a restricted neighbouring set of brain points, 
with a probability decreasing with the distance. 
%
  \item $\alpha(n+1)$ is determined adaptively by bisection in such a way that $p_{n+1}$ is close enough (\@but not too much, to avoid getting stuck\@) to $p_{n}$; the distance between the 
  two distributions is measured by means of the ratio $\tfrac{ESS\/(n+1)}{\plus60 ESS\/(n)}$ where the Effective Sample Size (\/ESS\/) is defined as 
  \begin{equation}
   ESS\/(n) =   \left[\ \sum_{i=1}^I\@ \left(w_n^\narc{i\@}\right)^2\ \right]^{-1}
  \end{equation}
and the weights are given by 
\begin{subequations}
 \begin{equation}
w_{n+1}^\narc{i\@}\ = \frac{\tilde w_{n+1}^\narc{i\@}}{\sum_{j=1}^I\tilde w_{n+1}^\narc{j\@} }\@ ,
\end{equation}
with
\begin{equation}
\tilde w_{n+1}^\narc{i\@}\ =\ w_{n}^\narc{i\@}\, \frac{p_{n+1}\big(X^\narc{i}_n\@|\@\mbf{y}\big)}{p_{n}\big(X^\narc{i}_n\@|\@\mbf{y}\big)}\,.
\end{equation}
\end{subequations}
The exponent $\alpha(n+1)$ is chosen in such a way that $\tfrac{ESS\/(n+1)}{\plus60 ESS\/(n)}$ falls between $0.9$ and $0.99$.

\item whenever the ESS falls below $\tfrac{I}{2}$, a systematic resampling step \cite{douc2005comparison} is applied in order to prevent all but one sample from having negligible weights.
\end{itemize}

\subsection{Estimates.}\label{SS2.4} 
The approximated posterior distribution contains information on multiple alternative models. 
In order to produce a sensible map, we first restrict our attention to the most probable model by estimating
the posterior probability for the number of sources, i.e. by computing 
\begin{equation}
\begin{split}
\hat{n}_D &= \arg\max\, \mathbb{P}(n_D\@|\@\mbf{y}\@) =\\
          &= \arg\max \left[\@\sum_{i=1}^I w^\narc{i} \delta\left(n_D^{},n_D^\narc{i\/}\right)\right]\,,
\label{eq:mod_sel}
\end{split}
\end{equation}
being $\delta(\@\cdot\@,\@\cdot\@)$ the Kronecker delta.

Subsequently, for each voxel $r$, we compute
\begin{equation}
\mathbb{P}\/(r\@|\@\mbf{y},\hat{n}_D) = \sum_{i=1}^I w^\narc{i} \delta\left(\hat{n}_D^{},n_D^\narc{i}\right) \sum_{k=1}^{n_D^{\nard{i}}} \delta\left(r, r_k^{\narc{i}}\right)\, ,
\label{eq:estimate}
\end{equation}
which represents the posterior probability of a dipole being located in $r$. In the analyses below this quantity is used to produce posterior maps of activation. In addition, we compute estimates of dipole locations as the local peaks of this probability map.
Finally, dipole moments can be reasonably estimated as the mean (\ref{eq:post_q}\@ a) of the corresponding Gaussian distribution. 

\section{Results}\label{S3}
In order to assess the performance of SASMC in the frequency domain source modeling, we first tested it against MEG synthetic data
and then we carried out the localization of \emph{post--movement beta rebound} (\/PMBR\/) activity from a go/no--go experimental dataset.
As a touchstone, in both cases we compared the results provided by SASMC with those given by DICS.

For a correct interpretation of the results given below, care must be taken of the different nature of the algorithms:
the images produced by SASMC represent the marginal posterior probability of source location (equation \eqref{eq:estimate}) 
and their spread has to be interpreted as localization uncertainty, while DICS maps show the ratio of source versus noise power.
%
Importantly, the representation of the results is therefore clearly affected by the setting of the visualization threshold. Owing to the
explained differences between the methods, it seems reasonable to use a different value for each method. In fact, we could use a single set of values for the posterior probability maps produced by the SASMC, namely Fmin = 1e-4, Fmid = 2e-3, Fmax = 5e-2; the lower bound Fmin is approximately the uniform value of the prior distribution: if the posterior is below this value, it indicates that the data are not increasing the posterior probability of that specific location. The visualization thresholds of the DICS maps, on the other hand, were tuned in order to avoid too widespread sources, or missing sources.

The prior parameters in the SASMC were fixed as described below. In particular, the noise standard deviation did not need tuning thanks to the presence of a pre--whitening step. The regularization parameter in DICS was optimized heuristically, in order to obtain the best possible images. The analysis with DICS has been performed using the MNE--Python package (\/v\@$0.15$\/) \cite{gramfort2014mne}.

The computational cost of the SASMC algorithm is highly variable, depending on the number of particles, on the complexity of the posterior distribution and on the estimated number of sources. In the simulations below, performed on a standard laptop (CPU Intel\textsuperscript{\textregistered} Core\textsuperscript{\texttrademark}
i$5$-$4210$U @ $1.7$\@GHz, RAM 8.00\@GB), the running time ranged between few minutes and few hours.

\subsection{Simulated Data}\label{SS3.1}

\medskip
\subsubsection{Synthetic data generation}\label{SSS3.1.1}

\smallskip
We devised two distinct synthetic scenarios.
In \textit{Scenario 1} data are produced by three dipolar sources; several time series are simulated, with varying signal--to--noise ratio (\/SNR\/) and intersource correlation level. In \textit{Scenario 2} the three dipoles are replaced by corresponding dipole clusters, mimicking the effect of extended sources. In order to make the two scenarios comparable, the strength of each current dipole of Scenario 1 is equally split into the strengths of the dipoles belonging to the corresponding cluster.

For both Scenarios syntetic data have been generated by means of a three--shell forward model and of a high resolution source space with $20453$ vertices, corresponding to an average spacing of 3 mm.
The geometry of the MEG device corresponds to that of a $306$ channels Elekta Neuromag\textsuperscript{\textregistered} Vector View system. 
The forward problem (\/computation of the lead field\/) has been solved by means of a boundary element method, as implemented in the MNE--Python package, starting from the geometry of the head of a real subject.

The source time courses are all $10$\@Hz sinusoids, each modulated by a Gaussian.
The sinusoids may differ by their phase, while the Gaussians have different means.

\begin{figure*}
\begin{center}
\begin{tikzpicture}
[every node/.style={minimum height={1.5cm},thick,align=center}]
\node[inner sep = 0pt](S1) at (0,0.5) {\includegraphics[width=4.5cm]{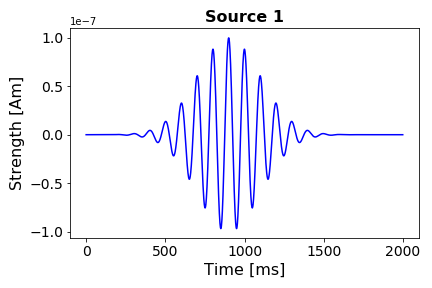}};
\node[inner sep = 0pt](S2) at (0,-2.5) {\includegraphics[width=4.5cm]{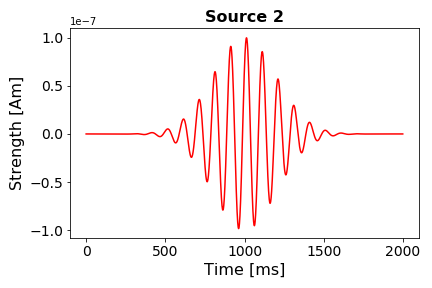}};
\node[inner sep = 0pt](S3) at (0,-5.6) {\includegraphics[width=4.5cm]{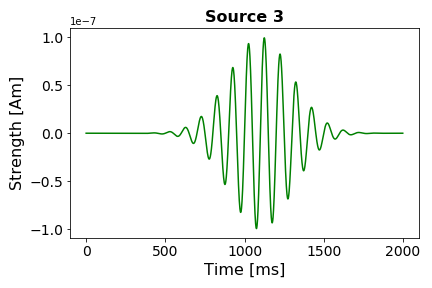}};
\node[inner sep = 0pt](NL) at (5.3,-2.55) {\includegraphics[width=4.7cm]{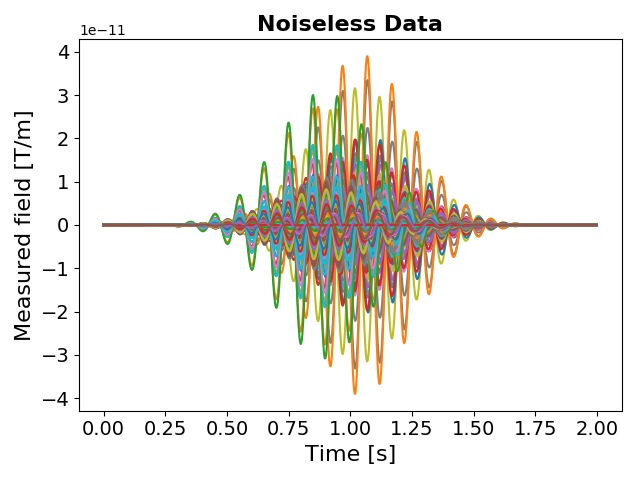}};
\node[inner sep = 0pt](N1) at (12,-0.5) {\includegraphics[width=4.7cm]{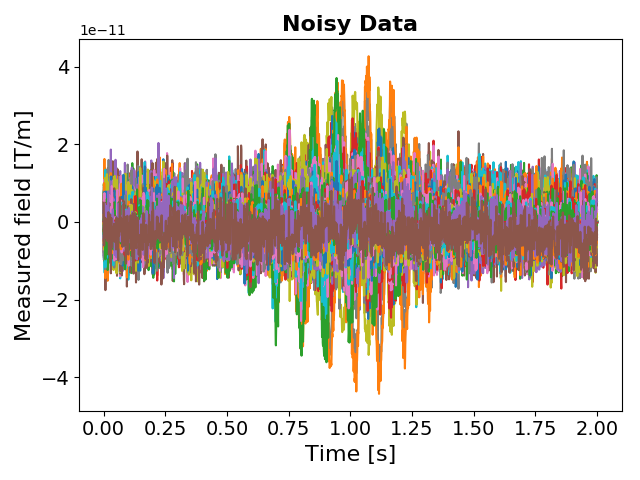}};
\node[inner sep = 0pt](N2) at (12,-5) {\includegraphics[width=4.7cm]{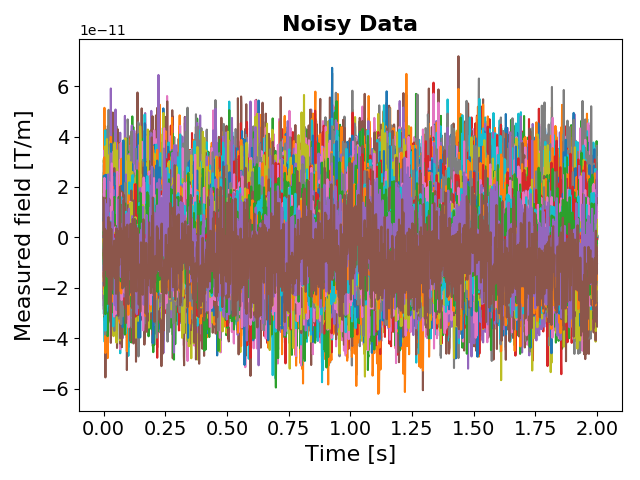}};
\draw[->] (S1) -- (NL);
\draw[->] (S2) -- (NL);
\draw[->] (S3) -- (NL);
\def\myshift#1{\raisebox{1ex}}
\draw[->,postaction={decorate,decoration={text along path,text align=center,text={|\sffamily\myshift|SNR = -5dB}}}] (NL) -- (N1) ;
\def\myshift#1{\raisebox{1ex}}
\draw[->,,postaction={decorate,decoration={text along path,text align=center,text={|\sffamily\myshift|SNR = -15dB}}}] (NL) -- (N2);
\end{tikzpicture}
\end{center}
\caption{Three source scenario\@: dipole time courses ($10$\@Hz sinusoids modulated by Gaussian windows with different means, $\Delta\@{\phi_{1\@2}} = \pi/4$, 
$\Delta\@{\phi_{1\@3}} = \pi/2$) are first combined to obtain noiseless data through the forward model; 
empty room noise is then added, with a given SNR.}
\label{fig:data}
\end{figure*}
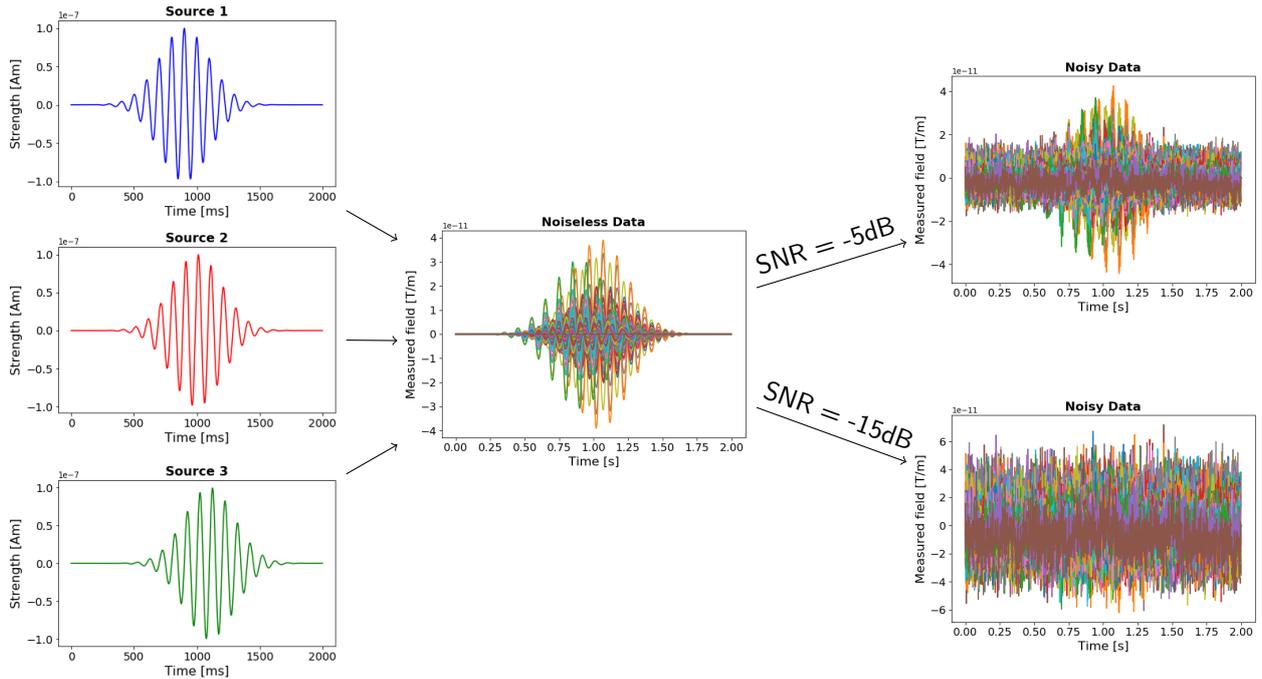

Each dataset consists in (the simulation of) a $88$ seconds long recording, corrupted by \@``\/empty room\@'' noise, obtained from an MEG recording without a subject. The temporal resolution of the recording is $1$\@ms.
SNR levels, measured in decibels as
\[
10\ \log_{10}\@\left( \frac{\sum_t \|\textbf{y}_t\|^2}{\sum_t \|\textbf{n}_t\|^2}\right)\, ,
\]
take values ranging from $0$\@dB to $-15$\@dB with a $5$\@dB step. 
At the lowest case, the signal happens to be completely buried in noise (\/Figure \ref{fig:data}, notice the different scale on the $y$ axis in the noisy data pictures\/).

\smallskip

\smallskip

All sources are situated in the left hemisphere and oriented along the $x$ axis in the Neuromag coordinate system. In Scenario 1, numbering the sources from front to back, the distance between source $1$ and source $2$ is $1.6$ cm, while the posterior source 3 is located $8.5$ cm away from source 2. In Scenario 2, three dipole clusters are grown around the three sources of Scenario 1, and include their nearest neighbours: the anterior and the middle cluster contain $4$ dipoles, while the posterior cluster contains $7$ dipoles; dipoles belonging to the same cluster have the same time course.

Denoting by $\Delta\@\phi_{i\/j}$ the phase difference between the $j$--th
and the $i$--th dipole time course, phase differences take the following values: 
$\Delta\@\phi_{1\/2} \in \{\/0, \pi/4, \pi/2\/\}$, $\Delta\@\phi_{1\/3} \in \{\/0, \pi/4, \pi/2\/\}$.

\subsubsection{Data preprocessing and inversion settings}

Each simulated recording has been segmented into $44$ non--overlapping epochs of $2000$ time--points.
A Hanning window has been applied to each epoch to reduce spectral leakage and then data have been Fourier transformed. 

Data in the frequency band from $9$ Hz to $11$ Hz have then been selected from all the epochs.

\smallskip
For the analysis with DICS, both data and noise cross--spectral density matrices have been computed. 

In the SASMC analysis pipeline, data underwent a pre--whitening step before being Fourier transformed.

Both the noise CSD matrix and the noise covariance matrix have been computed starting from an empty room recording that was not 
used in the data generation process.

To avoid inverse crime, both the source grid and the lead field matrix used by the inversion methods are different
from those used to generate the data. In particular, the forward model is now single--shell while the source grid is defined by a different decimation of the white matter surface which comprises only $8181$ vertices, corresponding to an average spacing of 5 mm. In the images of the following sections, we will be using blue points to indicate those vertices of the coarser source grid which are nearest to the true source locations. We notice that this also affects the number of dipoles in the dipole clusters of Scenario 2, which becomes $3$ for the anterior and middle cluster, and $4$ for the posterior cluster.

\subsubsection{Scenario 1}\label{sec:sim_1}
\smallskip

%
%

\begin{figure}[htb]
\centering
\subfigure{\includegraphics[width=4cm]{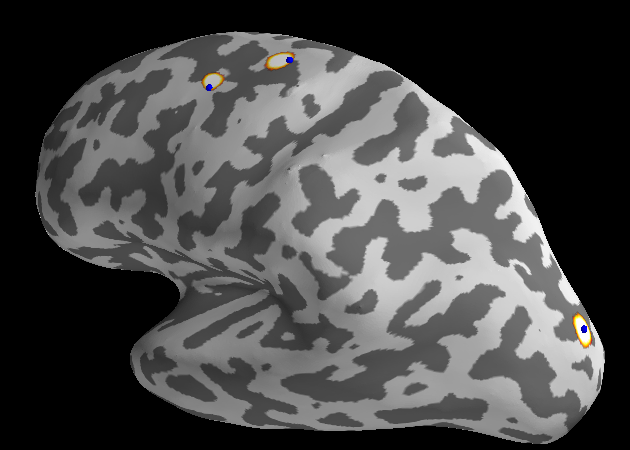}}
\subfigure{\includegraphics[width=4cm]{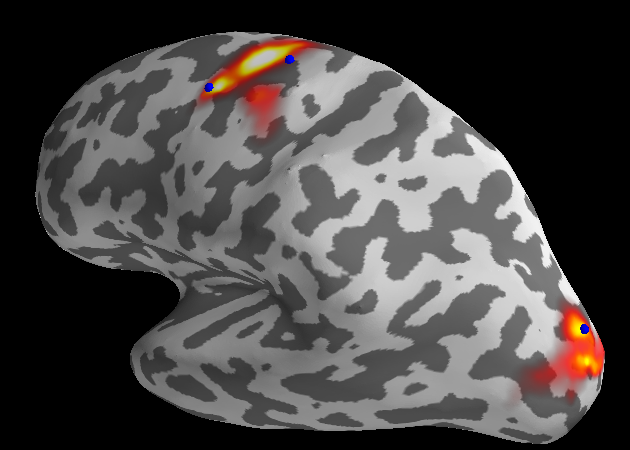}}\\[-3pt]
\subfigure{\includegraphics[width=4cm]{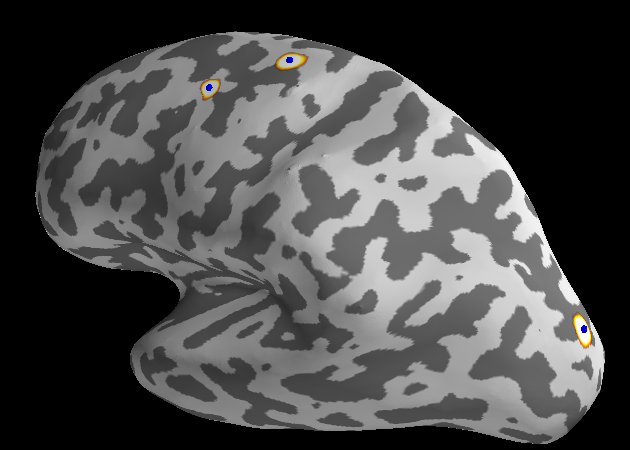}}
\subfigure{\includegraphics[width=4cm]{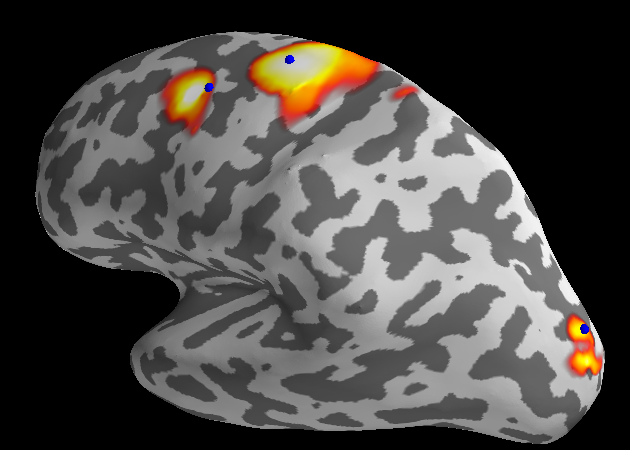}}\\[-3pt]
\subfigure{\includegraphics[width=4cm]{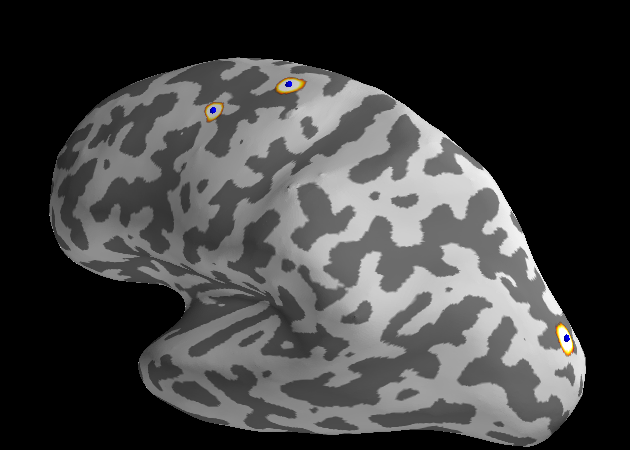}}
\subfigure{\includegraphics[width=4cm]{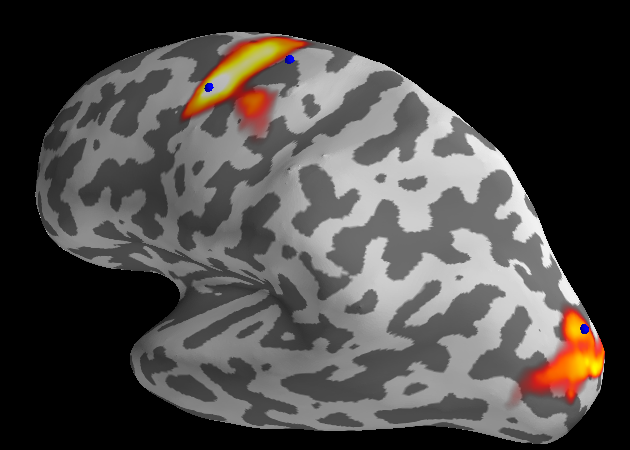}}
\caption{Scenario 1. Results at SNR = $-5$\@dB of the three dipoles syntetic scenario. 
         SASMC (\/left column\/) and DICS (\/right column\/) results are shown as color maps;  true source locations are represented as blue spots.           
         Top row: results for $\Delta\@\phi_{1\/2} = \pi/4$, $\Delta\@\phi_{1\/3} = \pi /2$; 
         Second row: results for $\Delta\@\phi_{1\/2} = \pi/2$, $\Delta\@\phi_{1\/3} = \pi /4$;
         Bottom row: results for $\Delta\@\phi_{1\/2} = \Delta\@\phi_{1\/3} = 0$.}
\label{fig:sim2_a}
\end{figure}

\begin{figure}[htb]
\subfigure{\includegraphics[width=4cm]{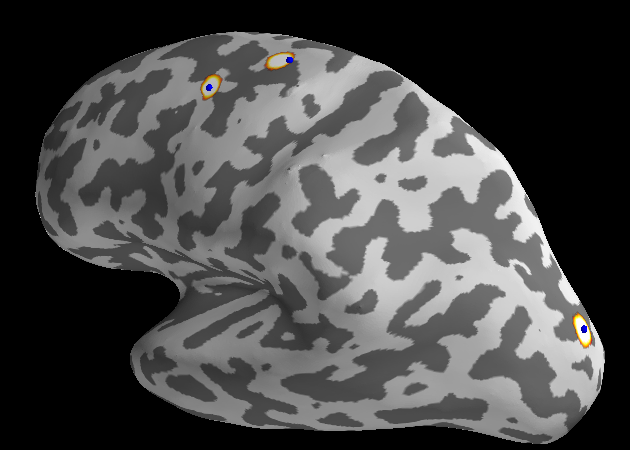}}
\subfigure{\includegraphics[width=4cm]{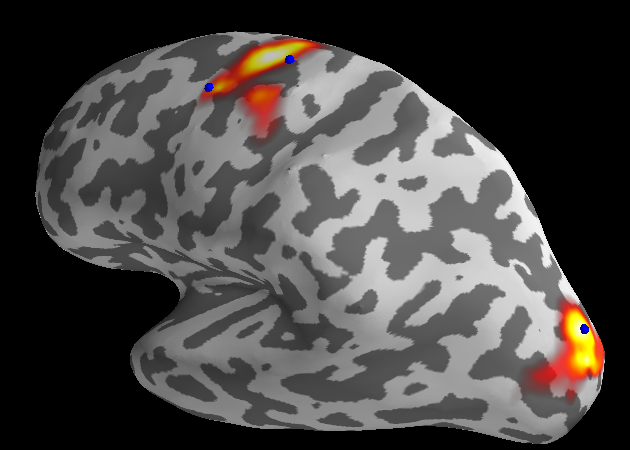}}\\[-3pt]
\subfigure{\includegraphics[width=4cm]{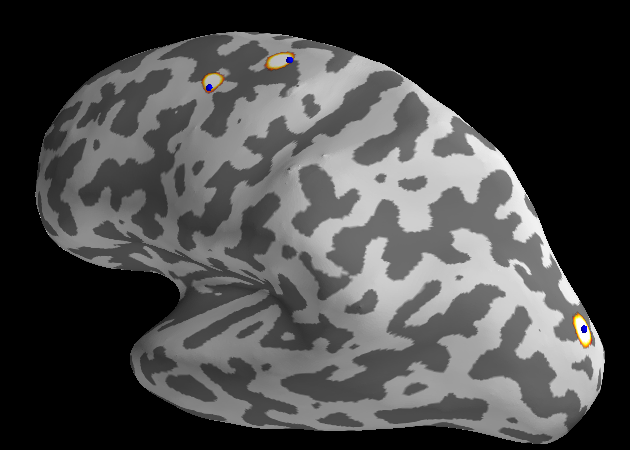}}
\subfigure{\includegraphics[width=4cm]{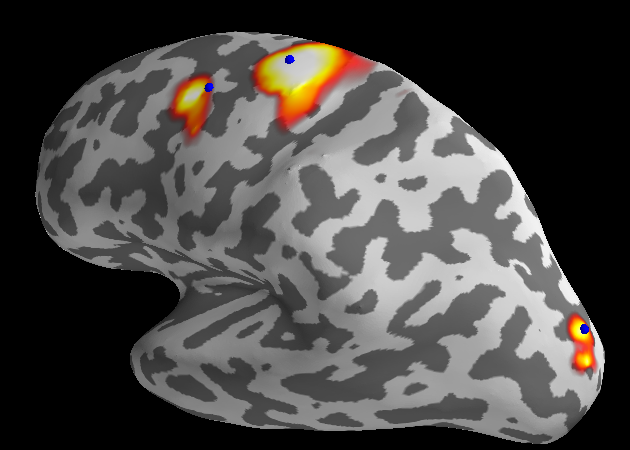}}\\[-3pt]
\subfigure{\includegraphics[width=4cm]{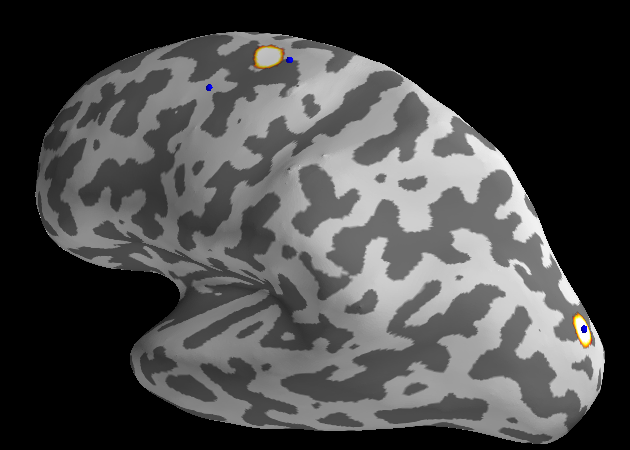}}
\subfigure{\includegraphics[width=4cm]{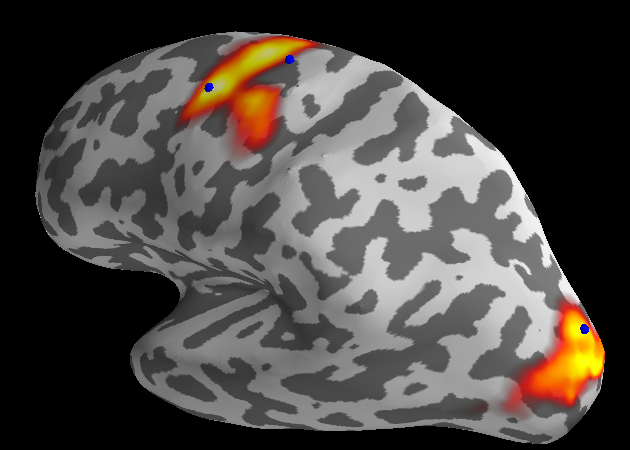}}
\caption{Scenario 1. Results at SNR =$\@-15$\@dB. 
         SASMC (\/left column\/) and DICS (\/right column\/) results are shown as color maps; source grid points nearest to true source locations are represented as blue spots.           
         Top row: results for $\Delta\@\phi_{1\/2} = \pi/4$, $\Delta\@\phi_{1\/3} = \pi /2$; 
         Second row: results for $\Delta\@\phi_{1\/2} = \pi/2$, $\Delta\@\phi_{1\/3} = \pi /4$;
         Bottom row: results for $\Delta\@\phi_{1\/2} = \Delta\@\phi_{1\/3} = 0$.}
\label{fig:sim2_b}
\end{figure}

In Figures \ref{fig:sim2_a} and \ref{fig:sim2_b} we report the results of the analysis of the three dipole scenario at SNR = $-5$\@dB and SNR=$-15$\@dB, respectively, and three different intersource correlation levels. The SASMC was used with $\lambda = 0.25$ and $\sigma_q=$1e-5 Ams; since the time series contain 2,000 time points, this corresponds to a prior standard deviation of about $10^{-8}$ Am in the time domain. 
Only the left lobe is portrayed since both methods correctly reconstruct no activity in the right hemisphere.

In five out of six cases, the posterior distribution approximated by the SASMC indicates a three--dipole model, and the localizations are correct; in the last case, the one with SNR=$-15$\@dB and zero phase difference between the three sources, the posterior distribution indicates a two--dipole model, the posterior dipole is localized correctly while the middle dipole is slightly mis--localized and the anterior dipole is lost. 
This is most likely due to the higher level of noise affecting the data, combined with the different split of the data between real and imaginary parts, due to the zero phase difference.

The activity maps provided by DICS appear to be in good agreement with the results of the SASMC and are not influenced by the different SNR values in the explored range; however, DICS reconstructions are affected by the intersource phase difference to a greater extent and, while the general picture is similar, in the first row and in the last row it is difficult to tell apart source 1 and source 2, while source 3 might as well be two nearby sources.

\begin{figure}
	\begin{center}
	\includegraphics[height=3.2cm]{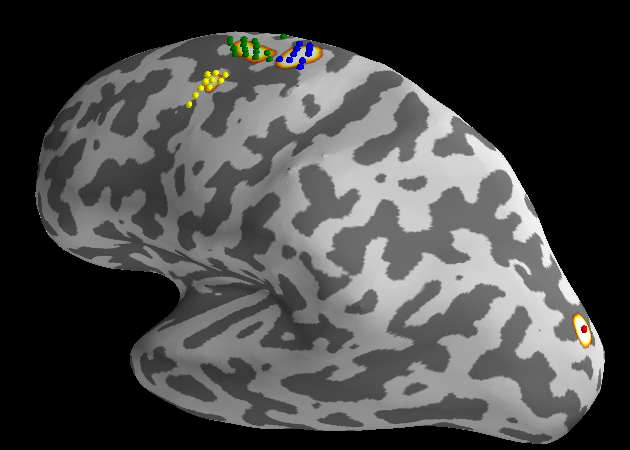}
	\includegraphics[height=3.2cm]{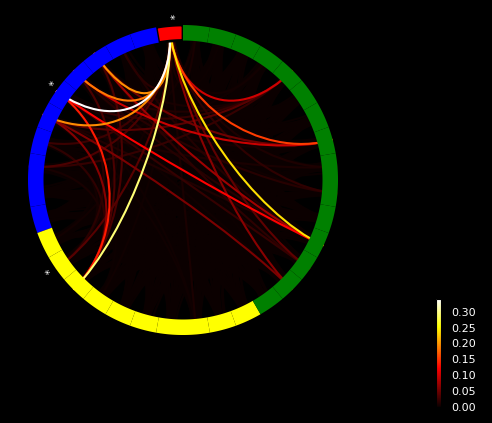}
	\end{center}
	\caption{Scenario 1. Left: posterior map obtained from a single trial; colored points represent the discrete support of the posterior distribution, clustered based on their belonging to different probability areas. Right: uncertainty quantification through dipole location correlation.}
	\label{fig:uncertainty}
\end{figure}

We notice that the posterior maps of the dipole locations are highly focused; this lack of uncertainty is due to the large amount of data that we are feeding the algorithm with; indeed, data from 44 epochs are used, and this provides a highly peaked posterior distribution.
In many cases, such large amount of data is not available; therefore we tried to apply the SASMC to the analysis of a single epoch.
In Figure \ref{fig:uncertainty} we show the results obtained by the SASMC applied to the analysis of a single epoch, in the case SNR = $-15 $ \@dB, $\Delta\@\phi_{1\/2} = \pi/2$, $\Delta\@\phi_{1\/3} = \pi /4$. In the left panel, the posterior distribution displays a non--negligible spread, corresponding to higher uncertainty particularly on the location of the anterior dipoles. In fact, four distinct high--probability regions appear in the map; however, the posterior distribution of the number of sources assignes more than $99\%$ probability to the three--dipole model, which indicates that two of these areas are alternative dipole locations. To investigate this point, in the right panel of Figure \ref{fig:uncertainty} we plot the correlation between dipole locations: essentially, we 
show how source locations are linked with each other in the Monte Carlo samples; location pairs appearing more often are linked by lighter lines in the plot. The plot shows that the red source is linked to a blue source and either to a green or to a yellow source; on the other hand, the yellow and the green area are never linked, indicating that sources belonging to these two areas are mutually exclusive. Eventually, the posterior distribution from the single trial correctly localizes the posterior source and the middle source (although with some uncertainty) and indicates a third and final source on either side of the superior--frontal sulcus, with higher probability close to the correct location. 

%

\subsubsection{Scenario 2}\label{sec:sim_2}

We now present the results obtained by the SASMC when analyzing the data generated by dipole clusters. 
We first notice that the statistical model described in Section \ref{SS2.2} allows freedom in choosing two parameters of the prior distribution: the expected value of the number of sources $\lambda$ and the expected source strength $\sigma_q$. When one explicitely aims at estimating a multi--dipole model, it is convenient to use the standard \textit{dipolar} setting already used in Scenario 1, in which  $\lambda = 0.25$ and $\sigma_q = 1$~e-5 Ams favour low--dimensional models. When, on the other hand, one expects extended sources, 
a different \textit{cluster} setting, in which the prior distribution gives higher probability to larger number of weaker sources by using $\lambda = 3$ and $\sigma_q=3$ e-6 Ams, can better represent our prior information.
We also notice that estimating the size of the active area from M/EEG data is known to be a hard problem; it is therefore interesting to investigate what happens when the distributed--source setting is applied to the data of Scenario 1, that have been generated by strictly dipolar sources.

\begin{figure}[htb]
\subfigure{\includegraphics[width=4cm]{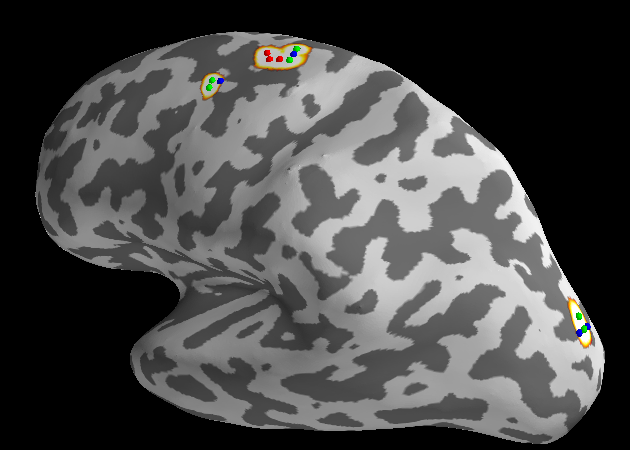}}
\subfigure{\includegraphics[width=4cm]{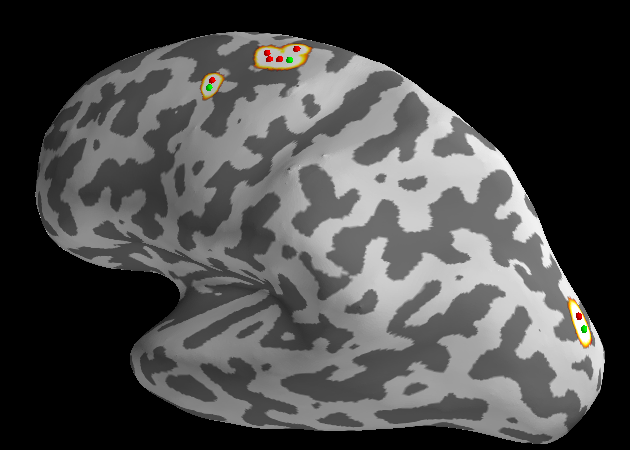}}\\[-3pt]
\subfigure{\includegraphics[width=4cm]{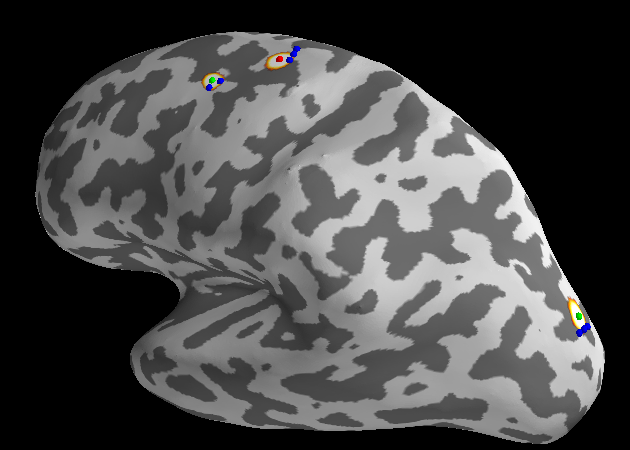}}
\subfigure{\includegraphics[width=4cm]{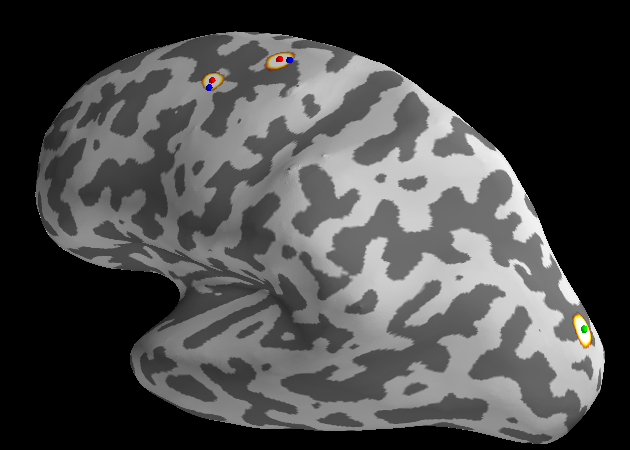}}\\
\caption{Scenario 2. SASMC posterior maps and point estimates. Left column: true sources are dipole clusters; right column: true sources are single dipoles. Top row: cluster setting used in the SASMC; bottom row: dipolar settings used. Color coding: blue points are true sources not reconstructed; green points are true sources correctly reconstructed; red points are estimated sources not corresponding to true sources.}
\label{fig:sim_clustered}
\end{figure}

In Figure \ref{fig:sim_clustered} we collect the results: in the left column we show the posterior distribution obtained by the SASMC from the data generated by the three extended sources, with the cluster setting (top panel) and with the dipolar setting (bottom panel). In the right column we show the posterior distribution obtained from the data generated by three dipolar sources, again with the cluster setting in the top panel and with the dipolar setting in the bottom panel. 
True source locations are plotted as blue points; estimated source locations are plotted as green points when they hit a true source location, and as red points otherwise.

The Figure shows that the SASMC correctly reconstructs activity in three distinct regions; however, the estimated number of sources -- which is related to the extent of the estimated source clusters -- depends on the prior parameters rather than on the true underlying source configuration. Indeed, when using the cluster setting, the anterior source is estimated as a two--dipole cluster, the middle source is estimated as a five--dipole cluster and the posterior source is estimated as a two--dipole cluster; this happens independently on whether the true underlying configuration is formed by three dipolar sources or three dipole clusters. Similarly, when using the dipolar setting, only three dipolar sources are estimated independently on the underlying true source configuration.

\begin{figure}[htb]
	\subfigure{\includegraphics[width=4cm]{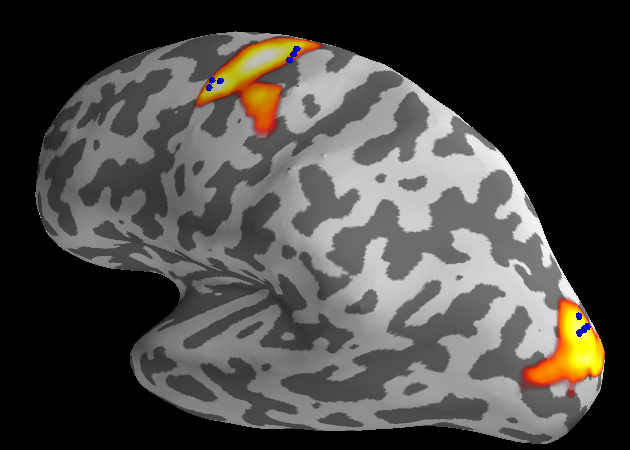}}
	\subfigure{\includegraphics[width=4cm]{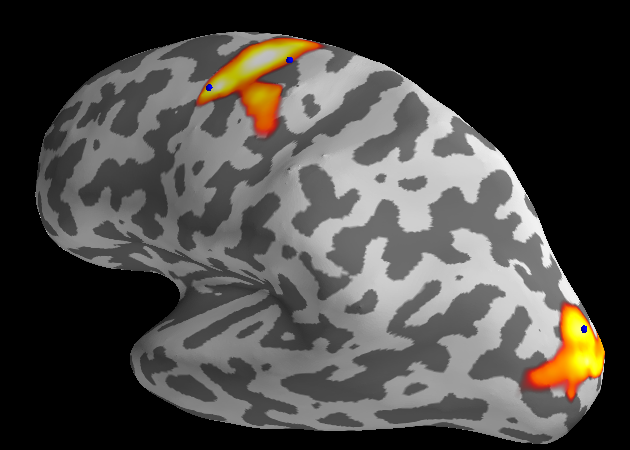}}
	
	\caption{DICS maps obtained from the data generated by the dipole clusters of Scenario 2 (left) and by the single dipoles of Scenario 1 (right).}
	\label{fig:DICS_distributed}
\end{figure}

We notice that the same behaviour affects the maps computed by DICS: in the left panel of Figure \ref{fig:DICS_distributed} we show the DICS map obtained from the data generated by three clusters, while in the right panel we plot again, for comparison, the DICS map obtained from the data generated by three single dipoles. The two maps are almost identical.

\subsection{Experimental Data}\label{SS3.2}

\medskip 
\subsubsection{Experimental design and MEG recordings}\label{SSS3.2.1}

\smallskip
The experiment consisted in a visual go/no--go task \cite{recagaen11}. 
Green and red circles, preceded by a fixation point, were presented at the center of a black screen and participants were instructed 
to respond to green (\/go\/) stimuli using their dominant hand and to keep still whenever a red circle (\/no--go\/) appeared.

The movement was a brisk extension of the hand, monitored by electromyography (\/EMG\/) and visual observation. 
Subjects were trained for several minutes in order to keep their head as still as possible and to endeavour to avoid blinking and mirror movements throughout the experiment.
The motor task consisted of $100$ go and $50$ no--go trials;
the duration of each stimulus was $500$ ms while inter--stimulus interval (\/ISI\/) was randomized between $5$ and $8$ seconds.

\smallskip
Neuromagnetic activity was recorded during the motor task using a whole head $306$--channel Neuromag  MEG system (\/Triux, Elekta Oy, Helsinki, Finland\/),
located in a two--layer magnetically shielded room with active shielding engaged.

Electroculogram (\/EOG\/) and electrocardiogram (\/ECG\/) recordings were also acquired, and used for artifact removal. 
EMG activity was obtained by means of pairs of Ag/AgCl surface electrodes placed bilaterally over the index and carp flexor muscles. 

The locations of the Head Position Indicator (\/HPI\/) coils, together with three anatomical landmarks (\/left/right auricular, nasion\/) and 
$600$--$800$ points on the skull scalp of the subjects, were determined with a three--dimensional digitizer (\/FastTrack, Polhemus, Colchester, USA\/) 
to allow alignment of the MEG and magnetic resonance image coordinate systems. 
The HPI coils were then maintained activated during the recordings throughout the whole experiment to monitor head movements.

All the data were recorded at the rate of $1$ kHz and online band--pass filtered (\/$0.03$--$330$ Hz\/).
Noise reduction was performed with the temporal extension of signal source separation \cite{tasi06} (t-SSS, MaxFilter Elekta Neuromag Oy).
T1 weighted MR images of the subjects brain were acquired using a 1.5\@T Siemens Avanto scanner or 3\@T Philips Achieva system. 
A realistic model of the cortex was obtained from T1 weighted MR Image by means of  Freesurfer \cite{fischl2012freesurfer} and 
then used to compute the forward model through a boundary element method.

\medskip
\subsubsection{Data preprocessing}\label{SSS3.2.2}

\smallskip
Data of a single experiment, in which one healthy right--handed subject took part, have been chosen 
among the entire experimental dataset. 

\begin{figure}[htb]
\centering
\subfigure{\includegraphics[width=6cm]{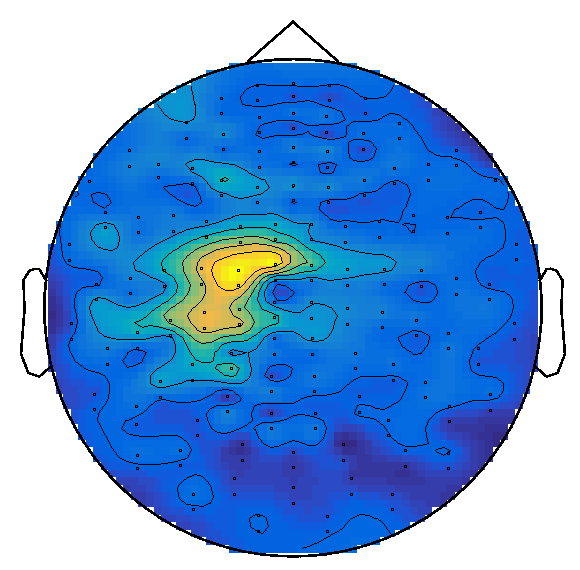}}\\[1pt]
\subfigure{\includegraphics[width=9cm]{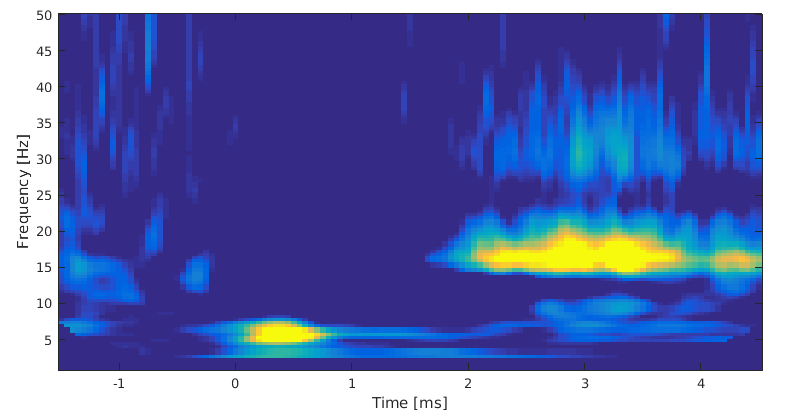}}
\caption{Top\@: Topographic map of the relative change of time--frequency compared to the baseline, in the selected time window (\/$2-4$s post stimulus\/) and 
in the beta frequency range (from $13.5$ to  $20.5$ Hz). 
Bottom\@: Mean time--frequency representation, normalized with pre--movement baseline data, of the sensors located in motor area contralateral to the movement. 
The figure represents the relative increase/decrease compared to the power in the baseline and clearly shows an increase in beta power in the selected area
after a movement.}
%
\label{fig:exp_prep}
\end{figure}

\smallskip
Several preprocessing steps have been carried out in order to prepare the data for the analysis.  
The preliminary operations consisted in:
\begin{itemize}
\item selection  of the artifact--free go stimuli epochs from $-2$\@s to $5$\@s with respect to the go stimulus presentation;
\item calculation of time--frequency representations (TFRs) of power over the selected trials. Figure \ref{fig:exp_prep} shows the mean TFR over 
      the sensors placed in the contralateral motor cortex;
\item definition of the time--frequency  window\@\footnote{\@PMBR frequency components, which in generic terms lie in the beta band, have been shown to 
be subject--specific \cite{pfurtscheller1999event}. } on the basis of the TFR above\@: specifically, we selected the time window after the subject's movement,
between $2$ and $4$ seconds after the trigger, and the frequency band from $13.5$ to  $20.5$ Hz;
\item application of the Hanning window to the artifact--free epochs, and Fourier transformation of the data;
\item evaluation of the SNR of each epoch, defined in the following way \cite{mahjoory2017consistency}: first, selection from each epoch of the single topography corresponding to the peak of the signal of the sensors placed in the controlateral motor cortex;
second, computation of the ratio between the spectral power
of the selected topography and the average spectral power in 2 Hz wide 
side bands to the left and to the right of the latter for each sensor in the controlateral motor cortex;
third, average among sensors;
\item selection of those epochs having a SNR greater or equal to $1.5\@$dB.
\end{itemize}

The above pre--processing steps resulted in $17$ epochs from which data in the frequency band $13.5-20.5$ Hz have been selected for the analysis.

\smallskip
Once again, the SASMC analysis pipeline provided for a pre--whitening step before applying the Fourier transform, while, for the analysis with 
DICS to be carried out, both the data and the noise CSD matrix were computed. In this real scenario, data from $0.5$\@s before the stimulus
to $1.5$\@s after the stimulus were considered as noise (see Figure \ref{fig:exp_prep}).


\subsubsection{Source modeling}\label{SSS3.2.3}

\begin{figure}[thb]
\centering
\subfigure{\includegraphics[width=4cm]{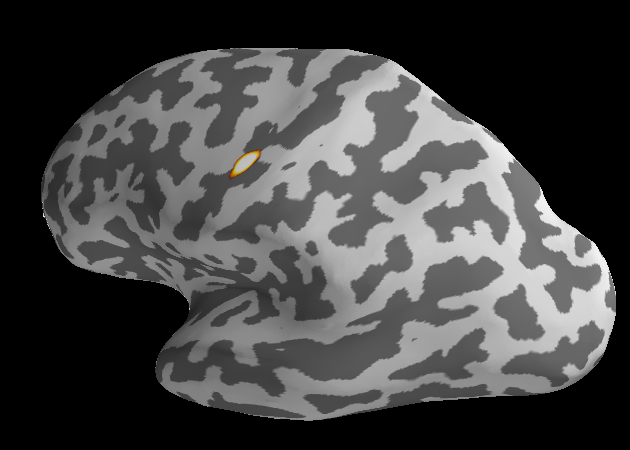}}
\subfigure{\includegraphics[width=4cm]{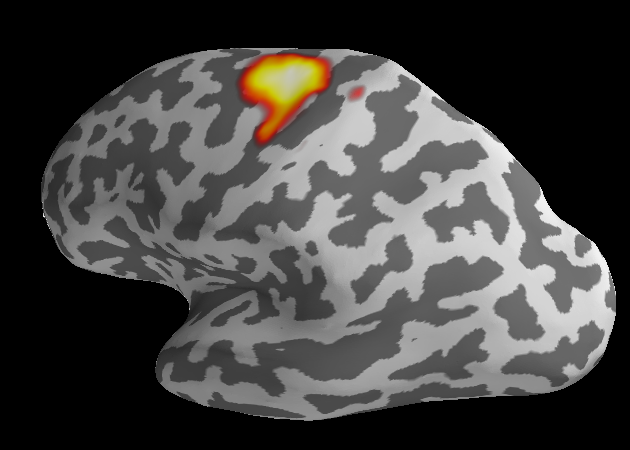}}\\[-3pt]
\subfigure{\includegraphics[width=4cm]{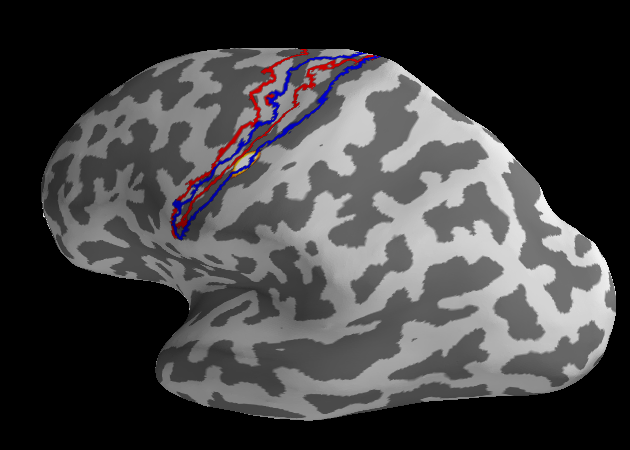}}
\subfigure{\includegraphics[width=4cm]{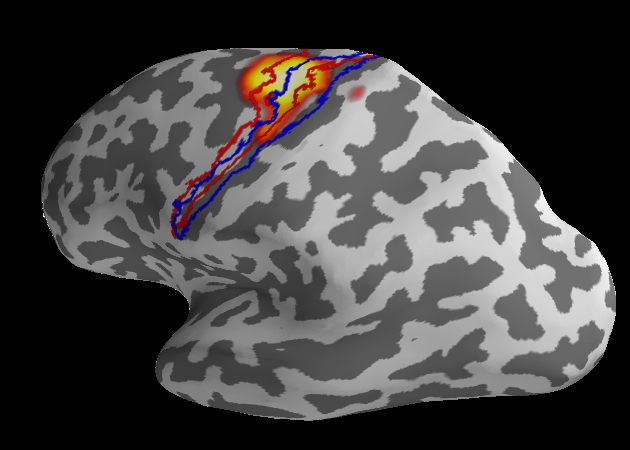}}
\caption{Analysis of the experimental dataset. SASMC (\/left column\/) and DICS (\/right column\/) results are shown as color maps. 
Bottom row: results with posterior primary motor Brodmann areas superimposed.}
\label{fig:exp_res}
\end{figure}

\smallskip
We applied the SASMC in the dipolar configuration. 
In Figure \ref{fig:exp_res} we show the posterior probability map obtained by the SASMC sampler, in the left column, and the map computed by DICS, in the right column; in the bottom line of the Figure, anterior (red) and posterior (blue) primary motor areas are depicted, as obtained by Freesurfer.

Analogously with what happened in the simulations, the posterior distribution approximated by the SASMC is more focused than the DICS map, and roughly consistent in terms of location. Specifically, the DICS map points to a more mesial area, where input from arm and shoulder is expected, while the SASMC localization is closer to the hand area. On the other hand, the posterior probability is mostly in the sulcus, while the DICS map covers the gyrus. Both localizations appear to be in accordance with the Brodmann area classification. Similar results have been obtained in \cite{cheyne2013meg,jurkiewicz2006post}.

\section{Discussion}\label{S4}


\smallskip
In this study we presented a Bayesian approach for localization of multiple dipolar sources in the frequency domain, in which the posterior distribution is approximated by a Sequential Monte Carlo sampler.
We systematically compared the results of the proposed method with those obtained by DICS, a well--known method for frequency domain source analysis. 

We first applied the method to simulated data, in which the underlying true source distribution is known. 
We designed a first synthetic scenario with three dipolar sources oscillating at the same frequency, whose time courses were modulated by Gaussian functions with different shifts. We let the inter--source phase difference and the signal--to--noise ratio vary in plausible intervals.
The results from the SASMC were consistently good across different conditions, with the only exception of the highest noise, fully correlated condition, in which one source was missing. The results of DICS appeared similarly good, to the extent that activity was detected in the surroundings of each true source; the DICS results do not appear to be influenced by the signal--to--noise ratio of the data, in the explored range; on the other hand, the DICS maps estimated from the same true source distribution at distinct inter--source phase differences appear more diverse than one would expect. Importantly, we also showed that the Bayesian approach presented here allows for uncertainty quantification, not only in terms of accuracy of individual source locations, but also in terms of cross--correlation between simultaneously active sources.

In a second synthetic scenario we explored the behaviour of the method when the true sources are not dipolar, which is a common condition when studying brain rhythms. We simulated three extended sources by using clusters of nearby dipoles. 
Here we showed that, when given proper prior information, the method is capable of reconstructing a relatively large number of dipoles in the correct locations, or close to them. In this sense, the method can go beyond the classical multi--dipole modeling in which each active area is represented by just a single dipole. However, if the same data set is analyzed with a prior distribution encouraging fewer and more intense sources, then the active areas are correctly localized, but the estimated configuration is just three dipoles. This suggested that the proposed method is robust to the presence of small active areas in terms of spatial localization but is not capable of inferring the source extent; indeed, the number of estimated dipoles depends on the prior distribution rather than on the data. To confirm this, we re--analyzed the data from the first scenario with the cluster settings, and observed that the reconstructed configuration was three dipole clusters rather than the three single dipoles. Indeed, it is well known that estimating the source extent from M/EEG data is a hard problem; the proposed approach does not seem to be suited to this task, in its current form. In this sense, DICS does not appear to perform better: in all scenarios DICS tends to reconstruct blurred activations around the true source locations, due to source leakage.

We finally applied the SASMC sampler for localizing the post--movement beta rebound in one healthy subject.
The results provided by the SASMC sampler in this case appear to be coherent with the literature on this topic, 
and also in good agreement with those provided by DICS. The SASMC localization seems to be closer to the hand area than that provided by DICS.

Our results indicate the SASMC sampler as an effective method for source localization in the frequency domain. While introduced for multi--dipole estimation, the method provides consistent results when less focal sources are sought. From a theoretical perspective, our model exploits the information on the phase of the signal (\/by using the Fourier Transform of the data\/) and does not use the data covariance matrix, whose rank is reduced by source correlation and can affect localization. In this sense, an improvement with respect to DICS can be expected. A major drawback of the SASMC is its computational cost, which can be significantly higher than that of DICS; parallelization of the computation is however feasible and should guarantee considerable performance improvements.


 
\section*{Acknowledgements}

AS was partially supported by Gruppo Nazionale per il Calcolo Scientifico (GNCS) -- INDAM.
SS was supported by the Aalto Brain Center (\texttt{http://brainscience.aalto.fi}).

\section*{References}
\bibliography{biblio}

\end{document}